\documentclass[12pt,a4paper]{article}

\usepackage{epsf}

\begin{document}

\begin{flushright}
  CPT--PC549--0797
\end{flushright}
\vspace{2\baselineskip}
\begin{center}
  \setcounter{footnote}{0}
  \renewcommand{\thefootnote}{\fnsymbol{footnote}}
  {\Large THEORY OF ELASTIC \\} \smallskip {\Large VECTOR MESON
    PRODUCTION: \\} \smallskip {\Large SOME RECENT DEVELOPMENTS
    \footnote{Talk given at the Ringberg Workshop on New Trends in
      HERA Physics, Ringberg Castle, Germany, 25--30 May 1997} \\}
  \vspace{2\baselineskip}
  \setcounter{footnote}{6}
  {\large Markus Diehl} \\ \medskip {\it
    CPT{\hspace{3pt}\footnote{Unit\'e propre 14 du Centre National de
        la Recherche Scientifique.}}, Ecole Polytechnique,
    91128 Palaiseau, France} \\
  \vspace{3\baselineskip}
  {\bf Abstract} \\
  \vspace{\baselineskip} \parbox{0.9\textwidth}{I review some recent
    developments in the QCD description of elastic vector meson
    production, focusing on issues like the meson wave function, the
    gluon density in the proton, factorisation, the use of
    parton-hadron duality and the nonperturbative vacuum.}
\end{center}
\vspace{\baselineskip}
\setcounter{footnote}{0}
\renewcommand{\thefootnote}{\alph{footnote}}

\section{Introduction}
I will discuss some recent developments in the description of elastic
vector meson production at high energy within QCD, restricting myself
to the cases where either the photon virtuality $Q^2$ or the vector
meson mass $M_V$ is large while the squared momentum transfer $t$ from
the proton is small. I will highlight progress and problems in the
theory of these processes, concerning issues such as the meson wave
function, the gluon density in the proton and factorisation. I shall
also spend some time on new approaches, namely the use of
parton-hadron duality~\cite{MRT} and a nonperturbative description of
the scattering~\cite{Dosch}. My aim is not to attempt a detailed
comparison of theory predictions with data but rather to indicate
where and why the predictions of various authors are different. I will
not have time to speak about phenomenological models based on Regge
theory.

\section{Simple approach}
\begin{figure}
  \vspace*{13pt} 
  \leftline{\hfill\vbox{\hrule width 0cm height0.001pt}\hfill}
  \begin{center} \leavevmode
    \setlength{\unitlength}{1cm}
    \begin{picture}(10,6)
      \put(2,0){\epsfxsize 6cm \epsfbox{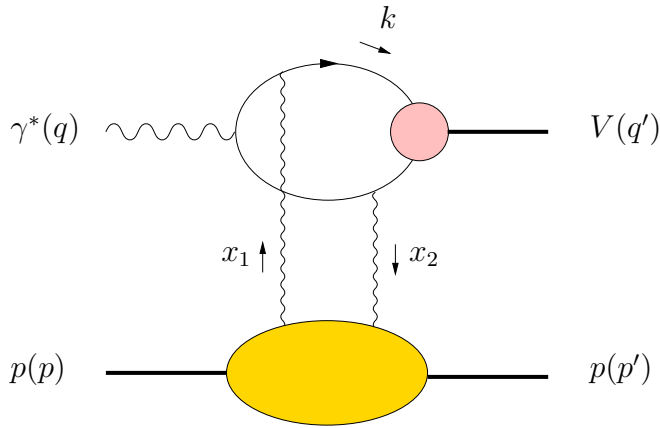}}
      \put(5.7,5.3){$k$}
      \put(0.8,0.65){$p(p)$}
      \put(0.8,3.85){$\gamma^\ast(q)$}
      \put(8.5,0.65){$p(p')$}
      \put(8.5,3.85){$V(q')$}
      \put(3.6,2.2){$x_1$}
      \put(6.1,2.2){$x_2$}
    \end{picture}
    \end{center}
  \leftline{\hfill\vbox{\hrule width 0cm height0.001pt}\hfill}
  \caption{One of the Feynman diagrams for $\gamma^\ast + p \to V +
    p$ with a vector meson $V = \rho$, $\rho'$, $\omega$, $\phi$, $J/
    \psi$, etc. More diagrams are obtained by attaching the gluons to
    the quark loop in different ways.}
  \label{fig:common}
\end{figure}

The common picture of most QCD models for elastic vector meson
production at high energy is shown in Fig.~\ref{fig:common}, where I
also define the kinematics. At large $\gamma^\ast p$ c.m.\ energy $W$
the dominating exchange is the pomeron, which is described by the
exchange of two gluons coupling on one side to the proton and on the
other to a quark loop to which are attached the virtual photon and the
vector meson. The vector meson couples to the quark line of the loop
by its wave function $\psi(z, k_T)$. Here $z$ and $k_T$ are defined by
a Sudakov decomposition for the quark momentum in the meson, $k = z q'
+ \zeta p + k_T$.

As a baseline let me present a very simple version of this model,
recently used by Cudell and Royen~\cite{CudRoy}. For the meson it takes
a nonrelativistic, constituent quark description, namely a wave
function peaked at $z = \frac{1}{2}$ and $k_T = 0$, so that the quark
and antiquark with mass $M_V /2$ share the four-momentum of the meson
equally. It makes the approximation that the exchanged gluons do not
interact and couple directly to the three constituent quarks of the
proton, following the work of Low, Nussinov, Gunion and
Soper~\cite{Oldies}.

With these assumptions Cudell and Royen find the $Q^2$- and
$t$-dependence of the cross sections for $\rho$, $\phi$ and $J/ \psi$
production in fair agreement with data from HERA, EMC and NMC, except
for problems with the fixed target $J/ \psi$ data. Let me remark that
the cross sections for longitudinal and transverse photons
respectively behave like $\sigma_L \sim 1 /Q^6$ and $\sigma_T \sim 1
/Q^8$ for $Q^2 \gg M_V^2$, while for nonasymptotic $Q^2$ the behaviour
in this variable is more complicated, even in this simple model where
a fixed strong coupling $\alpha_s$ was taken. One finds however a
ratio $\sigma_L / \sigma_T = Q^2 / M_V^2$ for all $Q^2 \ge 0$, which
is in disagreement with the data---a point I will come back to in
Sec.~4. As the approximation of noninteracting gluons does not allow
to describe the energy dependence of the cross section the authors
introduce an overall phenomenological factor in $\sigma_{L}$ and
$\sigma_{T}$, and find consistency with the data if this factor is
allowed to depend on $W$, but not on $Q^2$ or $M_V$. Little change in
the results is obtained when the propagators of the exchanged gluons
are taken as nonperturbative, as was done in earlier work of
Donnachie, Landshoff and Cudell~\cite{DLC}. Finally, the authors
observe that their model agrees with the data down to $Q^2 = 0$ even
for the light mesons $\rho$ and $\phi$. Notice that in this case one
has no large virtuality in the quark loop to justify the perturbative
treatment of the quarks and their coupling to the photon.

\section{Sophistication of the model}
Having seen that even the simplest form of the model depicted in
Fig.~\ref{fig:common} does not fare badly in describing data we will
now take a look at several of its refinements.

\subsection{Meson wave function}
The meson wave function $\psi(z, k_T)$ turns out to be a major source
of uncertainty in the predictions of the model. It determines which
virtualities dominate the integration over the quark loop and in
particular influences the overall normalisation of the cross section,
its $Q^2$-dependence and the production ratios for different mesons.
Various choices have been made in the
literature~\cite{Dosch,RRML,FSpsi,FSrho,NikZak,Hood,HalZhi}, sometimes
leading to appreciable differences in numerical results. For $J/ \psi$
photoproduction, as an example, Frankfurt et al.~\cite{FSpsi} find a
suppression of the cross section due to finite $k_T$ in the wave
function which is significantly stronger than the one estimated by
Ryskin et al.~\cite{RRML}. For $\rho$ electroproduction a wide range
of suppression factors has been obtained by Frankfurt et
al.~\cite{FSrho} according to the choice of $\psi(z, k_T)$, with a
particular sensitivity to its large-$k_T$ tail.

On the theory side it is known that as the renormalisation scale for
the wave function tends to infinity, corresponding to very large $Q^2$
in the process, the $k_T$-integrated wave function $\int d k_T^2 \,
\psi(z, k_T)$ behaves like $z (1 - z)$ due to QCD evolution, both for
longitudinal and transverse meson polarisation. This is very different
from the $\delta( z - \frac{1}{2})$ in the nonrelativistic wave
function, but the difficult question is of course to assess how far
one is from the asymptotic regime for the values of $Q^2$ one has in
experiment. Investigations of the wave function using the operator
product expansion and QCD sum rules have been made by Ball and
Braun~\cite{Ball} and Halperin and Zhitnitski~\cite{HalZhi}. The latter
find that at $z \to 0$ and $z \to 1$ the wave function should only
depend on $\frac{k_T^2}{z (1 - z)}$, which in particular excludes a
factorising ansatz $\psi(z, k_T) = \phi(z) \cdot \chi(k_T)$. They also
argue that from a conceptual point of view there should be no
perturbative tail like $1 / k_T^2$ in the wave function and that the
corresponding corrections to the meson-quark vertex should be
explicitly treated as $\alpha_s$-corrections to the leading order
result.

In the operator product expansion framework the effects of finite
$k_T$ in $\psi(z, k_T)$ are of higher twist. It is important to
realise that there are other higher twist contributions from diagrams
where one or more gluons enter the wave function blob in
Fig.~\ref{fig:common}. They correspond to higher fock states of the
meson such as $q\bar{q} g$ and $q\bar{q} g g$, which have respectively
been considered by Halperin~\cite{Halperin} and Hoodbhoy~\cite{Hood}.
Moreover, taking only into account a $q \bar{q}$ wave function
$\psi(z, k_T)$ is not gauge invariant. To see this consider that a
finite $k_T$ involves the transverse component of the operator $i
\partial_\mu$ between quark fields in a matrix element. To recover
gauge invariance one must pass to the covariant derivative, $i
\partial_\mu + A_\mu$, whose transverse component involves physical
gluon degrees of freedom.

Finally let me mention a problem raised by Frankfurt et
al.~\cite{FSpsi} concerning the $J/ \psi$ wave function. Calculating
the transition amplitude of a $J/ \psi$ to a timelike photon via a
charm quark loop with wave functions $\psi(z, k_T)$ obtained in
various nonrelativistic potential models they find that the region of
$k_T$ larger than the charm mass contributes up to 30\% of the loop
integral, thus calling into question the consistency of a
nonrelativistic description. Note that the transition $J/ \psi \to
\gamma^\ast \to e^+ e^-$ is commonly used to fix the normalisation of
the wave function from the semileptonic $J/ \psi$ decay width.

\subsection{Quark mass}
Which quark mass should be chosen in the loop of Fig.~\ref{fig:common}
is another source of uncertainty. Surprisingly this even holds for
heavy mesons. Ryskin et al.~\cite{RRML} point out that the cross
section scales like the eighth power of $M_{J/ \psi} / (2 m_c)$,
resulting in huge changes of its overall normalisation even if one
allows for a modest change in the charm mass $m_c$. Incidentally,
Frankfurt et al.~\cite{FSpsi} advocate to use the running quark mass
rather than the pole or constituent mass.

In the case of light quarks both the choice of current and constituent
masses has been made in the literature~\cite{RRML,FSrho,DLC,NikZak}.
Considering the limit of zero quark mass $m_q$ reveals that light
meson production from transverse photons is a higher twist
effect~\cite{RRML}: the cross section $\sigma_T$ is proportional to
$m_q^2$ in the collinear approximation for the quarks, i.e.\ if one
sets their transverse momentum $k_T$ to zero in the calculation of the
loop and uses the $k_T$-integrated meson wave function, corresponding
to what one does when using collinear parton densities in DIS. At
least one of the quantities $m_q$ or $k_T$ must be nonzero to obtain a
nonvanishing $\sigma_T$ from diagrams as shown in
Fig.~\ref{fig:common}.

\subsection{Off-diagonal gluon density}
We now focus on the lower blob in Fig.~\ref{fig:common}, the coupling
of the two gluons to the proton, including their interaction. If the
outgoing proton had the same momentum as the incoming one this blob
would be given by the gluon density in the proton. In our reaction
this is however not the case: even if the proton has zero transverse
momentum after the scattering it has lost a fraction $x = (M_V^2 +
Q^2) / (W^2 + Q^2)$ of its longitudinal momentum, which is necessary
to make a timelike vector meson out of a spacelike photon. Thus the
longitudinal momentum fractions $x_1$ and $x_2$ of the gluons with
respect to $p$ (cf.\ Fig.~\ref{fig:common}) are not equal, their
difference being $x_1 - x_2 = x$. One can argue that in the leading
$\ln Q^2 \cdot \ln (1 /x)$ approximation it is indeed the gluon
density $g(x)$ that describes the blob~\cite{All,Bart}, as the typical
values of $x_1$ and $x_2$ in the loop integration are of order $x$ and
to leading $\ln (1 /x)$ all such values are equivalent.

The most prominent phenomenological consequence of this description is
the strong rise of the cross section in $W$ when either of the scales
$Q^2$ or $M_V^2$ is hard. At this point I should remark that only the
imaginary part of the $\gamma^\ast p \to V p$ amplitude is actually
calculated from the Feynman diagrams of Fig.~\ref{fig:common}, cutting
them in the $s$-channel. The contribution of the real part to the
cross section is negligible if the $W$-dependence is weak as in the
case of soft pomeron exchange, otherwise it can be accounted for in an
approximate way~\cite{All,RRML}.

Beyond leading $\ln (1 /x)$ the quantity describing the blob in
Fig.~\ref{fig:common} is certainly different from the gluon
distribution that contributes to inclusive DIS, and there has been
much theoretical interest recently in such so-called ``off-diagonal
parton densities''. Strictly speaking they are not ``densities'' by
the way, since they correspond to matrix elements of parton fields
between \emph{different} proton states and do not have a probability
interpretation. Let me note that their evolution equations differ from
the DGLAP equations by their splitting
functions~\cite{BalBrauJi,RadyushFS}. Some studies have been
performed~\cite{Hood,RadyushFS} to estimate the difference between the
off-diagonal and the usual, diagonal gluon densities, finding no big
difference at small $x$, but much remains to be done in this field. An
obvious remark is that the dependence of the off-diagonal gluon
density on $t$ is specific of its asymmetric kinematics and cannot be
predicted by approximating it with the diagonal density measured in
other processes.

In the same way as in DIS at small $x$ one can also go beyond the
leading $\ln Q^2$ approximation and instead of $g(x)$ consider the
unintegrated gluon density $f(x,l_T)$ with a finite transverse
momentum $l_T$ of the gluons~\cite{RRML}. 

The question of which factorisation scale is appropriate in the gluon
density is numerically very important since we know from DIS that
parton distributions at small $x$ change rapidly with this scale.
Generically it is set by the typical virtualities in the quark loop,
i.e.\ by $Q^2$ and the quark mass, but various concrete choices have
been advocated in the literature: for $J/ \psi$ production the scale
$\frac{1}{4} (Q^2 + M_{J/ \psi}^2)$ used by Ryskin et al.~\cite{RRML}
is smaller than that of Frankfurt et al.~\cite{FSpsi}, whereas Nemchik
et al.~\cite{NikZak} choose $Q^2 + M_V^2$ times a factor between 0.07
and 0.2 depending on the meson and its polarisation. Let me however
emphasise that \emph{any} choice of factorisation scale can only be an
educated guess of which value will make higher order corrections
small. As in other reactions such as for instance jet production
experiment should feel free to try different scales $\textit{const}
\cdot (Q^2 + M_V^2)$, and also $\textit{const} \cdot Q^2$ for light
mesons.

\subsection{Factorisation theorem}
\begin{figure}
  \vspace*{13pt}
  \leftline{\hfill\vbox{\hrule width 0cm height0.001pt}\hfill}
  \begin{center}
    \setlength{\unitlength}{1cm}
    \begin{picture}(11.5,5)
      \put(2,0){\epsfxsize 7.5cm \epsfbox{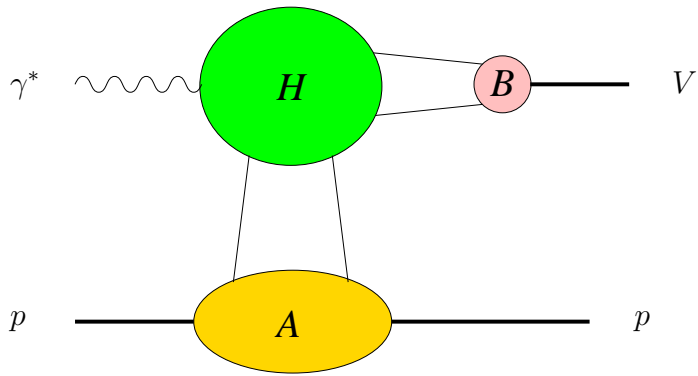}}
      \put(1.2,0.65){$p$}
      \put(1.2,3.75){$\gamma^\ast$}
      \put(9.5,0.65){$p$}
      \put(10.0,3.75){$V$}
    \end{picture}
    \end{center}
    \leftline{\hfill\vbox{\hrule width 0cm height0.001pt}\hfill}
    \caption{Factorisation of the amplitude for $\gamma^\ast p \to V
      p$ with a longitudinal photon into a hard scattering part $H$,
      an off-diagonal quark or gluon density $A$ and a meson wave
      function $B$.}
  \label{fig:factorise}
\end{figure}

For the asymptotic limit where $Q^2$ is much larger than all masses in
the process a factorisation theorem has been proven by Collins et
al.~\cite{Collins} within perturbative QCD. By an analysis of Feynman
diagrams and power counting arguments they find that to all orders in
$\alpha_s$ the amplitude for $\gamma^\ast p \to V p$ factorises into a
hard scattering part, a collinear, off-diagonal quark or gluon
distribution in the proton, and a collinear $q \bar{q}$ wave function
for the vector meson (see Fig.~\ref{fig:factorise}). The corresponding
amplitude has a power behaviour like $1 /Q$.

This power behaviour and factorisation property holds only for a
\emph{longitudinal photon}. If the $\gamma^\ast$ is transverse the
authors find that the amplitude only behaves like $1 /Q^2$, and that
there is no factorisation like in Fig.~\ref{fig:factorise}: soft,
``wee'' partons can bypass the hard scattering and directly go from
the proton blob $A$ to the meson blob $B$. The power behaviour in $Q$
confirms from a different point of view the statement that light meson
production from transverse photons is a higher twist process.

It is worth noting that these findings are not restricted to the
small-$x$ region but valid for general $x$, where the quark
distribution in the proton is relevant in addition to that of the
gluons.

Finally it is clear that the direct phenomenological application of
these results requires ``sufficiently large'' values of $Q^2$, and we
have already seen that higher twist effects such as the transverse
momentum in the meson wave function can be numerically important in
the HERA regime. Also one should keep in mind that the $Q^2$-behaviour
obtained by power counting can be strongly masked by logarithmic
scaling violations which are strong at small $x$.

\subsection{Rescattering effects}
Let me just briefly mention that several
papers~\cite{RRML,FSpsi,NikZak,GLM} have investigated the effects of
rescattering of the $q\bar{q}$-pair on the proton, in other words
shadowing corrections or multiple-pomeron exchange, and found that
they can be important in kinematic situations accessible at HERA.  An
observable that is naturally sensitive to such effects is the
$t$-dependence of the cross section, which has been investigated in
detail by Gotsman et al.~\cite{GLM}.

In this context one may remark that an exponential parametrisation $d
\sigma / d t \propto \exp( b t)$ at small and moderate values of $t$
is no more than a fit to a simple function: from theory one does not
expect such a behaviour to be exact. Remember for instance that
ordinary elastic form factors of hadrons are not exponentials. One
should be aware that just comparing an experimentally fitted slope
parameter $b$ with the logarithmic slope of $d \sigma / d t$ at $t =
0$ calculated in theory can be misleading. To compare the full
$t$-dependence is of course the most thorough way to proceed, but if
one looks for convenient ``handy'' parameters other choices than $b$
might be useful, such as the mean value $\langle t \rangle$ proposed
by Cudell and Royen~\cite{CudRoy}.

\section{The ratio $R = \sigma_L / \sigma_T$}
As I have already emphasised the physics of light meson production is
quite different with transverse and longitudinal photon polarisation.
It turns out that $\sigma_T$ is more sensitive to small quark
virtualities and thus to infrared physics than its counterpart for
longitudinal photons~\cite{MRT,FSrho,NikZak}. This can be understood
from the properties of the transition $\gamma^\ast \to q\bar{q} \,$:
while for a longitudinal photon configurations are preferred where the
longitudinal momenta of quark and antiquark are comparable, a
transverse photon likes to split into a $q\bar{q}$-pair where the
quark or antiquark carries only a small fraction of the photon
momentum and is soft. This is the aligned jet configuration, which is
of prime importance for the physics of diffraction. The interaction
with the gluons hardly changes the longitudinal momenta of $q$ and
$\bar{q}$, so that this configuration corresponds to small $z$ or $1 -
z$ in the meson wave function. To which extent $\sigma_T$ is dominated
by the soft region thus depends on how strongly small $z$ or $1 - z$
are suppressed in $\psi(z, k_T)$.

In the simple model presented in Sec.~2, where a meson wave function
peaked at $z = \frac{1}{2}$ is used, the ratio $R$ comes out as $Q^2 /
M_V^2$, which is far too big to describe the data for $Q^2$ in the
range of some ${\rm GeV}^2$. In the model of Nemchik et
al.~\cite{NikZak} the greater infrared sensitivity of $\sigma_T$ is
reflected in a lower factorisation scale of the gluon density than in
$\sigma_L$, and the authors find that $R$ grows less fast than
linearly in $Q^2$ and that its rate of growth in $Q^2$ depends on $x$.

\subsection{Using parton-hadron duality}
To circumvent the uncertainties associated with the meson wave
function in this delicate context, Martin et al.~\cite{MRT} invoke
parton-hadron duality: to obtain the cross section for
$\rho$-production they calculate the diffractive production of an open
$q \bar{q}$-pair by two-gluon exchange, project out the appropriate $q
\bar{q}$ partial wave to make a vector meson of given helicity, and
integrate the corresponding cross section over the $q \bar{q}$
invariant mass $M_{q \bar{q}}$ in an interval around $M_\rho$. The
authors stress that in this mass range partonic final states other
than $q \bar{q}$ are heavily suppressed and that the dominating
hadronic final state is a pair of pions. In the ratio $R$ several
uncertainties are expected to cancel to a large extent, e.g.\ 
uncertainties about the appropriate interval of $M_{q \bar{q}}$ and
about radiative corrections (the authors estimate that there should be
a large $K$-factor). As in the calculation by Nemchik et
al.~\cite{NikZak} $\sigma_T$ is dominated by lower quark virtualities
than $\sigma_L$, now in the context of open $q\bar{q}$-production, and
through the different factorisation scales in the gluon density the
authors obtain a ratio $R$ that grows less than linearly in $Q^2$ and
is in fair agreement with HERA data for $R$ in the range $5 {\ \rm
  GeV}^2 < Q^2 < 20 {\ \rm GeV}^2$. Again this ratio is predicted to
depend on $x$.

\section{A nonperturbative model}
To finish let me present an approach from a perspective of
nonperturbative physics. The model of Dosch et al.~\cite{Dosch} starts
from the high-energy scattering of a quark or antiquark, which is
described as the scattering in an external gluon field~\cite{Nacht},
like in the semiclassical model of diffraction of Buchm\"uller and
Hebecker~\cite{Heb}. This gluon field is then averaged over in the
sense of a path integral, according to the stochastic model of the
nonperturbative QCD vacuum~\cite{DoschVac}.  High-energy scattering
thus becomes related with nonperturbative parameters such as the SVZ
gluon condensate.

\begin{figure}
\vspace*{13pt}
\leftline{\hfill\vbox{\hrule width 0cm height0.001pt}\hfill}
  \begin{center}
    \setlength{\unitlength}{1cm}
    \begin{picture}(11.5,4.3)
      \put(2,0){\epsfxsize 7.5cm \epsfbox{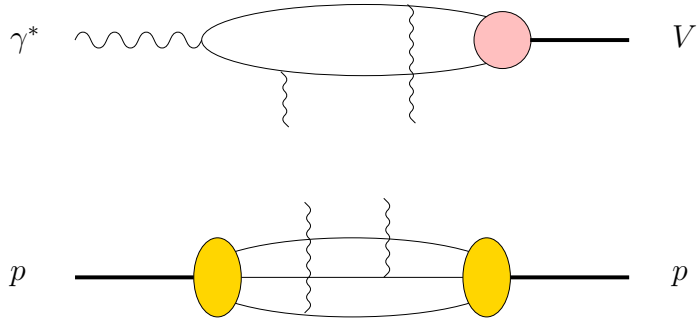}}
      \put(1.2,0.5){$p$}
      \put(1.2,3.6){$\gamma^\ast$}
      \put(10.0,0.5){$p$}
      \put(10.0,3.6){$V$}
    \end{picture}
    \end{center}
\leftline{\hfill\vbox{\hrule width 0cm height0.001pt}\hfill}
\caption{The process $\gamma^\ast p \to V p$ in the model of Dosch et
  al.~\cite{Dosch}. Quarks and antiquarks are scattered in the gluon
  field of the nonperturbative QCD vacuum.}
\label{fig:Dosch}
\end{figure}

The nonabelian character of the gluon field is crucial in this model:
not single quarks or antiquarks are scattered but rather colour
singlet configurations of $q\bar{q}$ or $qqq$, their transverse
separation having a strong influence on the strength of the
scattering. The physical picture for vector meson production is then
as shown in Fig.~\ref{fig:Dosch}. For the three-quark wave function of
the proton and the $q\bar{q}$ wave function of the vector meson an
ansatz has to be made, whereas the splitting of the virtual photon
into $q\bar{q}$ is calculated perturbatively. As in the simple
two-gluon exchange model discussed in Sec.~2 the $W$-dependence of the
cross section cannot be predicted in the present form of the model.

Its authors find fair agreement of their results with EMC and NMC data
for $Q^2$ between 2 and $10 {\ \rm GeV}^2$ and $W$ between 10 and 20
GeV, looking at observables such as the $Q^2$- and $t$-dependence of
the cross section and its normalisation, except that their cross
section for the $\phi$ is about a factor 2 above the data. They also
describe the ratio $R$ for $\rho$-production measured by NMC and the
$t$-dependence of $J/ \psi$ photoproduction.  They conclude by
emphasising the need to incorporate nonperturbative aspects of the
photon wave function for small $Q^2$ and $M_V$, and perturbative gluon
contributions which they expect to become important as $W$ increases
from fixed target to HERA energies.

\section{Conclusions}
The model depicted in Fig.~\ref{fig:common} is a good candidate for
the description of elastic vector meson production at large $Q^2$ or
meson mass in QCD. Even a very simple version of this model as
presented in Sec.~2 can describe quantitative features of the data.

Progress has been made in the theory of meson wave functions and of
the off-diagonal gluon distribution in the proton. For the asymptotic
regime of very large $Q^2$ a factorisation theorem has been worked out
which further elucidates the different physics of $\sigma_L$ and
$\sigma_T$.

On the phenomenological side various difficulties or uncertainties
present themselves if one is to make quantitative predictions in the
nonasymptotic regime, in particular for light meson production from
transverse photons, or if in turn one aims to extract information from
the data on the meson wave function or the gluon distribution. An
important task is to identify observables which are sensitive to only
a few theoretical effects.

Finally, as I showed in Sec.~4.1 and 5, there are promising
alternative approaches, which highlight the connection of this
reaction with open $q\bar{q}$-production, and with nonperturbative
physics and the QCD vacuum.

\section*{Acknowledgements}
It is a pleasure to acknowledge conversations with T. Gousset, R.
Klanner, P. V. Landshoff and A. D. Martin. Special thanks go to T.
Gousset for carefully reading the manuscript. Finally, I would like to
thank B. Kniehl, G. Kramer and A. Wagner for organising this very
fruitful workshop.

\end{document}